\documentclass[aps,prl,twocolumn,showpacs,amsmath,floatfix]{revtex4}
\pdfoutput=1 
\usepackage{amsmath,epsfig}
\usepackage[caption=false]{subfig}
\usepackage[active]{srcltx}
\usepackage{nicefrac}

\usepackage{array}
\usepackage{epstopdf}

\begin{document}
\title{{\it Ab initio} studies of ionization potentials of hydrated hydroxide and hydronium}
\author{Charles W. Swartz and Xifan Wu}
\affiliation{Department of Physics,  
Temple University, Philadelphia, Pennsylvania 19122, USA}

\begin{abstract}
The ionization potential distributions of hydrated hydroxide and hydronium are 
computed with many-body approach for electron excitations with 
configurations generated by {\it ab initio} molecular dynamics. 
The experimental features are well reproduced and found to be closely related to the 
molecular excitations. In the stable configurations, the ionization potential
is mainly perturbed by water molecules within the first solvation shell. 
On the other hand, electron excitation is delocalized on both proton receiving and donating 
complex during proton transfer, which shifts the excitation energies and broadens the spectra for both hydrated ions. 
\end{abstract}
\maketitle

The nature of the solvation structures of hydroxide (OH$^-$) and hydronium (H$_3$O$^+$) aqueous solutions
are of fundamental interest. It is the prerequisite to understand the mechanism of proton transfer (PT)
through the autoprolysis process in water, which is behind 
diverse phenomena in physics, chemistry and 
biology~\cite{Tuckerman_review_2010, Tuckerman_ACR, Parrinello_PNAS_2011, 
Hynes_Nature, Asthagiri_PNAS, JACS_exp, Chandra_PRL_2007}.
Photoemission spectroscopy (PES) has recently emerged as an important
experimental technique in elucidating the interactions between hydrated ions 
and surrounding water molecules~\cite{JACS_exp,Winter_JACS_2005}.
In PES experiment, the ionization potential (IP) 
is measured by the energy required to remove an electron from the molecule 
with respective to the Fermi level.
The state-of-art PES measurement is now able to detect the 
spectra signals from solvated ions which has been successfully used to determine
the IPs of hydrated OH$^-$ and H$_3$O$^+$~\cite{JACS_exp}.
To have an insightful understanding of the experiments, there is a critical need 
for theoretical modeling, at atomic scale,
which can unambiguously connect the IP and its distributions to its solvation 
structures.

Given the disordered structures of the liquids can be simulated by
{\it ab initio} molecular dynamics (AIMD)~\cite{AIMD,Patrick_PRB}, the PES spectra 
can be computed by the electron excitation
theory such as Hedin's GW self-energy approximation~\cite{Hedin, Onida}.
However, such a theoretical approach has not yet been applied to the study of 
the PES spectra in ion solutions. The difficulty lies in the fact that
the quasiparticle method scales unfavorably with the system size~\cite{Onida}.
On the other hand, a proper simulation of disordered liquid structure requires a large
supercell modeling. The computational burden is even more severe when it is necessary to 
take into account the statistical fluctuations of solvation structures 
both from the underlying H-bond network and more 
drastically, from the structural diffusion of PT. 
As a compromise, the computationally efficient static density functional
theory (DFT)~\cite{Tuckerman_LDA} or the semiclassical approximation was used~\cite{JACS_exp}. However, 
the static DFT designed for ground state electron minimization strongly underestimates the 
IP as an electron excitation property~\cite{Onida,Tuckerman_LDA,JACS_exp}.
Because of the same difficulty, the precise assignment of IPs for these 
hydrated ions has yet not been accurately determined.

To address the above issues, in the current work we generate the liquid 
structures of solvated hydroxide and hydronium in water by AIMD simulations 
and then compute the PES spectra within many-body formalism for electron
excitation. In particular, we adopt a recently implemented methodology treating 
the inhomogeneous electronic screening of the 
medium~\cite{Model_GW_1988, Wu_PRB_2009, Wei_PRL, Lingzhu_PRB_2012}, in which the 
maximally localized Wannier functions~\cite{MLWF} are used to greatly increase the 
computational efficiency. The IPs of hydrated hydronium and hydroxide
are determined by the real-space projection of quasiparticle density of states
on these solvated ion complexes. 
The resulting IPs and their distribution are in quantitative agreement with
experiments in both position and broadening.
The IPs of hydrated ions in liquid 
solutions are associated with the molecular excitation,
however, strongly influenced by the water molecules in the first solvation shell.
During the PT, the structural diffusion results in a delocalzied defect eigenstate.
As a result, the IPs of hydrated ions are broadened and shift into the 
main feature of bulk water. To the best of our knowledge, 
this is the first time that GW based quasiparticle theory has been applied to the PES spectra
of ion solutions. This approach will be also useful to study other important liquids.

Ion solutions are generated by two equilibrated AIMD trajectories~\cite{AIMD} containing either one
hydroxide or hydronium with 63 surrounding water molecules. A periodic cubic cell corresponding to
the experimental ambient density of water is used. Both simulations are preceded by a 3~ps
equilibrium run and then continued over a 25~ps time scale within the canonical ensemble. An elevated
temperature of ${\rm T}=330 {\rm K}$ is used, which has been found to approximately capture 
structural softening of H-bond in liquid water due to the quantum nuclear effect~\cite{Zhaofeng_Li}.
The atomic force is calculated by DFT using
the PBE functional~\cite{PBE} with a kinetic energy
cutoff of $70$ Ry.  Electron excitation calculations are completed using the static Coulomb hole and screened
exchange (COHSEX) approach combined with electron screening effects from the inhomogeneous medium
within the Hybertsen-Louie ansatz. All calculation are performed using the
QUANTUM-ESPRESSO code package~\cite{QuantumEspresso}.

\begin{figure}
   \captionsetup[subfigure]{labelformat=empty}
   \centering
   \subfloat[]{ \includegraphics[scale=0.55]{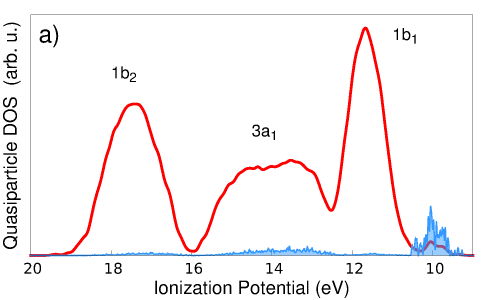}}
   \vspace{-0.5cm}
   \subfloat[]{ \includegraphics[scale=0.55]{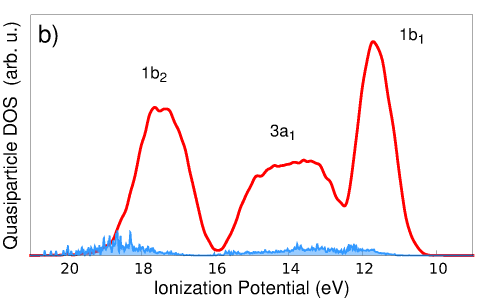}}
   \vspace{-0.6cm}
   \caption[]{ 
      (color online) Theoretical PES spectra (\textit{red line}) for the (a) OH$^-$ and (b) H$_3$O$^+$ ion solutions.
      The shaded (\textit{blue}) area indicates the IPs of hydrated OH$^-$ and H$_3$O$^+$~\cite{defect_id}.
   }
   \label{fig:DOS_lDOS}
\end{figure}

\begin{figure}
   
   \begin{tabular}{|m{1.5cm}|c|}

          \hline
          \centering
          H$_2$O &  \begin{tabular}{ c@{\hspace{2cm}}c@{\hspace{1.8cm}}c} $1b_2$ &$3a_1$ & $1b_1$\\ \end{tabular}\\
   
          \hline
          &\\
          Gas Phase & \begin{tabular}{c@{\hspace{0.380cm}}c@{\hspace{0.38cm}}c}\includegraphics[scale=0.03]{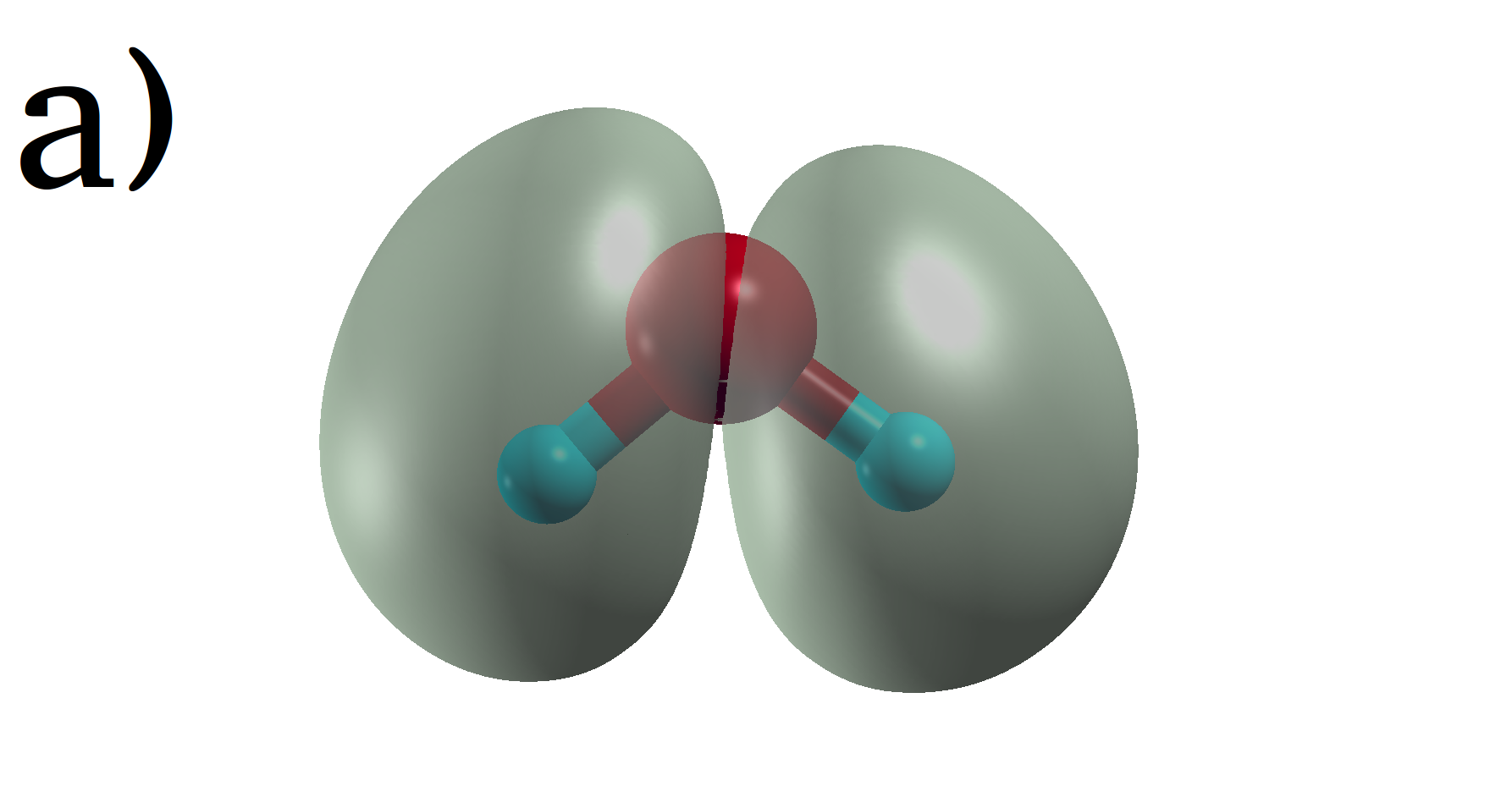} &\includegraphics[scale=0.03]{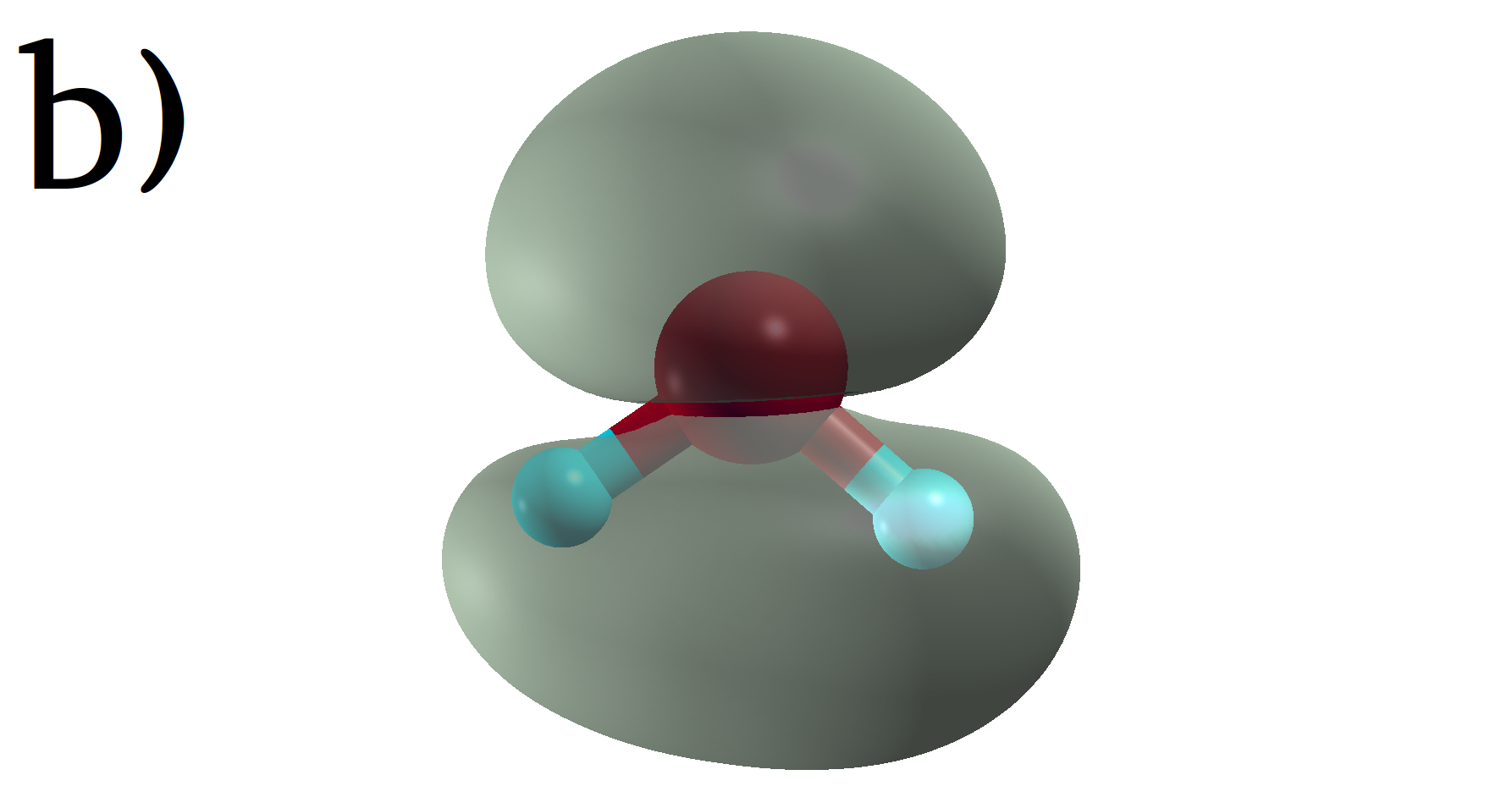} &\includegraphics[scale=0.03]{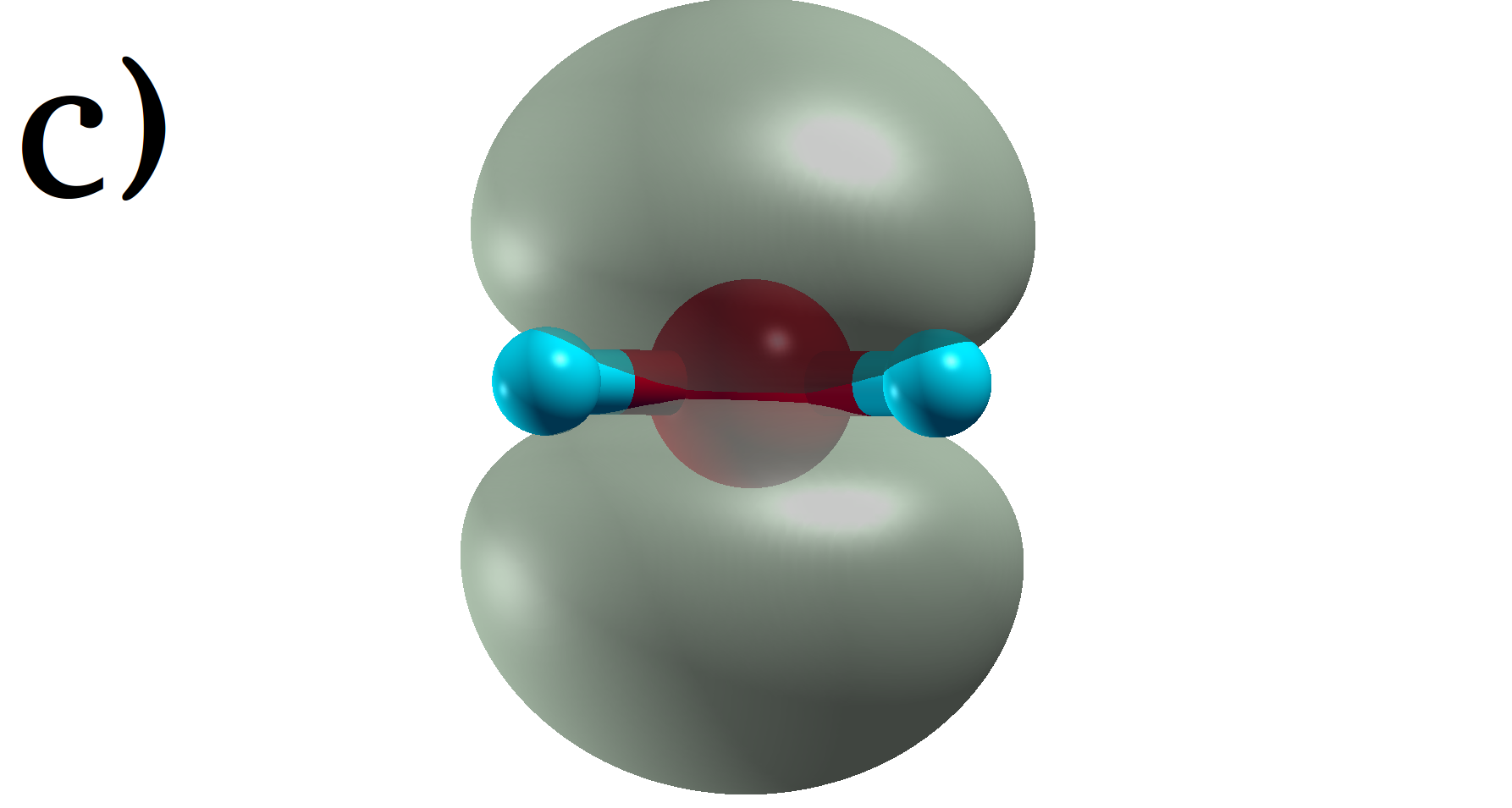}\\ \end{tabular}\\

          \hline
    \end{tabular}

   \begin{tabular}{|m{1.5cm}|c|}

          \hline
          \centering
          OH$^-$ & \begin{tabular}{ c@{\hspace{2.5cm}}c} $3a_1$ &$1b_1$\\ \end{tabular}\\
          \hline

          &\\
          Gas Phase & \begin{tabular}[c]{c@{\hspace{1.955cm}}c}  \includegraphics[scale=0.03]{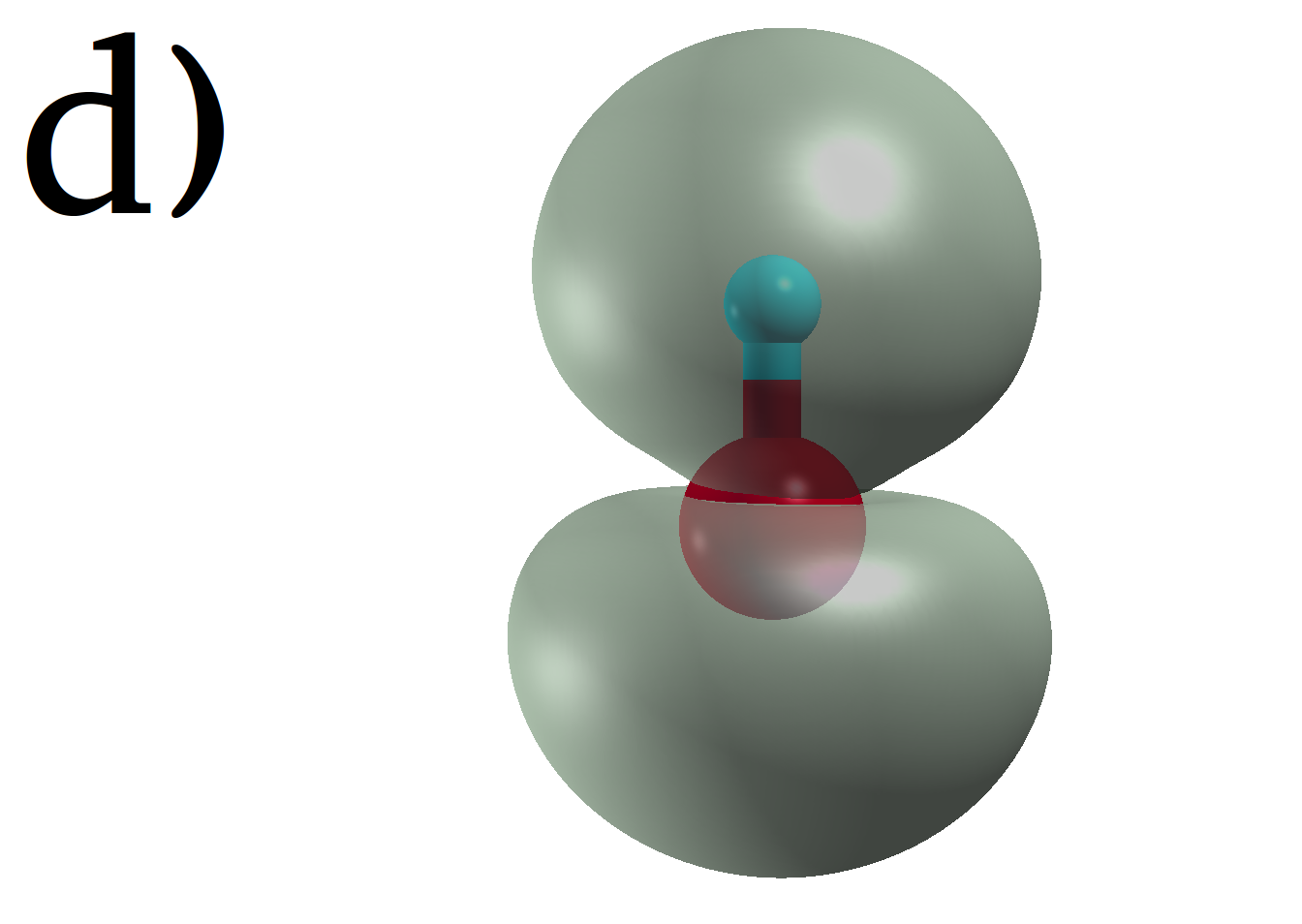} &\includegraphics[scale=0.03]{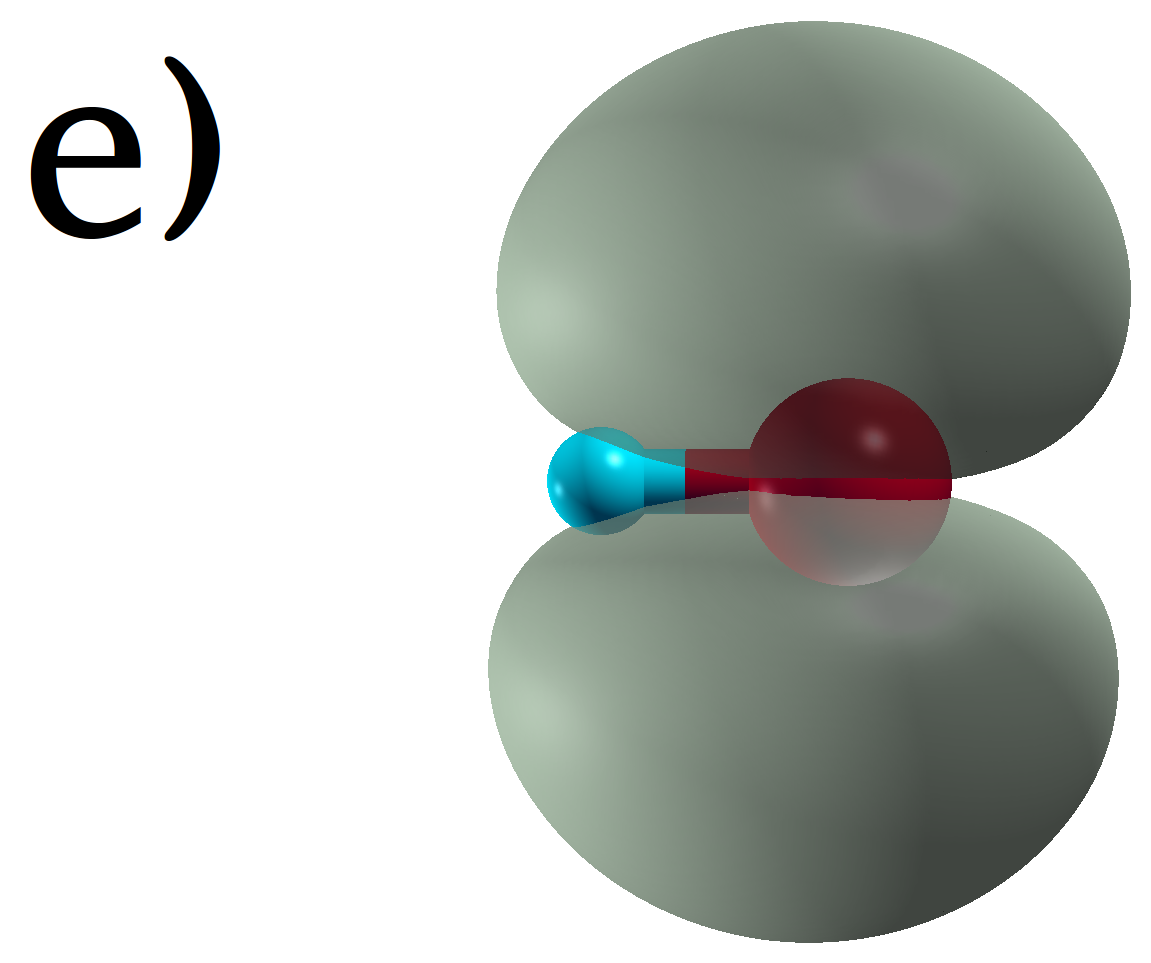}\\ \end{tabular}\\

          Liquid Solution: \newline No PT & \begin{tabular}[c]{cc}  \includegraphics[scale=0.0748]{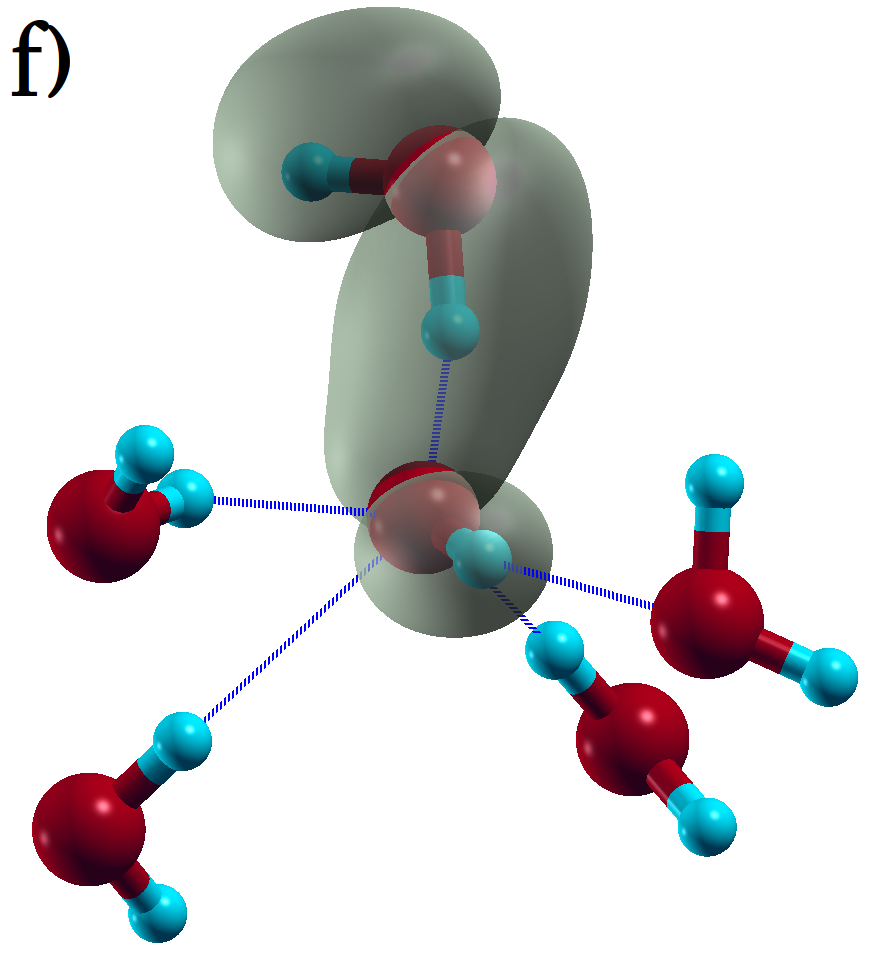} &\includegraphics[scale=0.075]{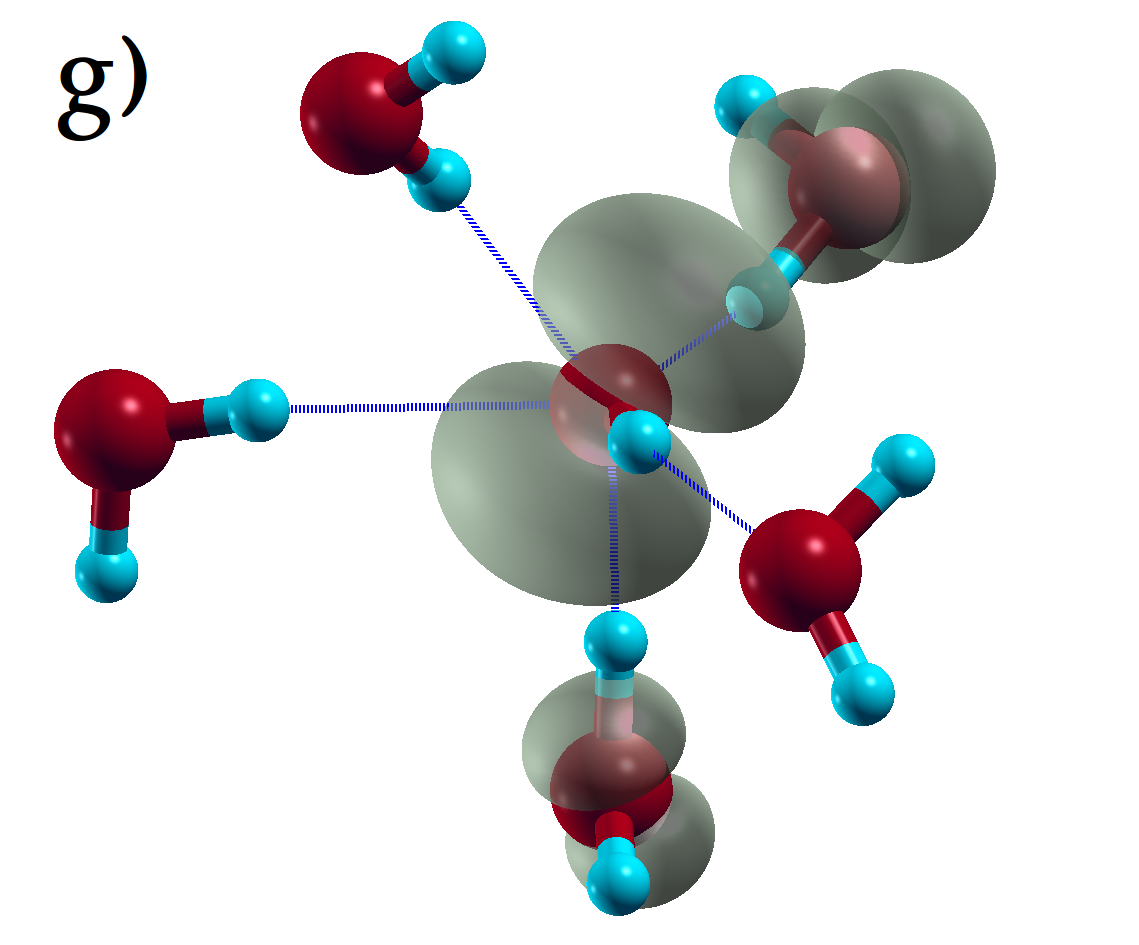}\\ \end{tabular}\\

          Liquid Solution: \newline PT & \begin{tabular}[c]{c@{\hspace{0.38cm}}c} \includegraphics[scale=0.09]{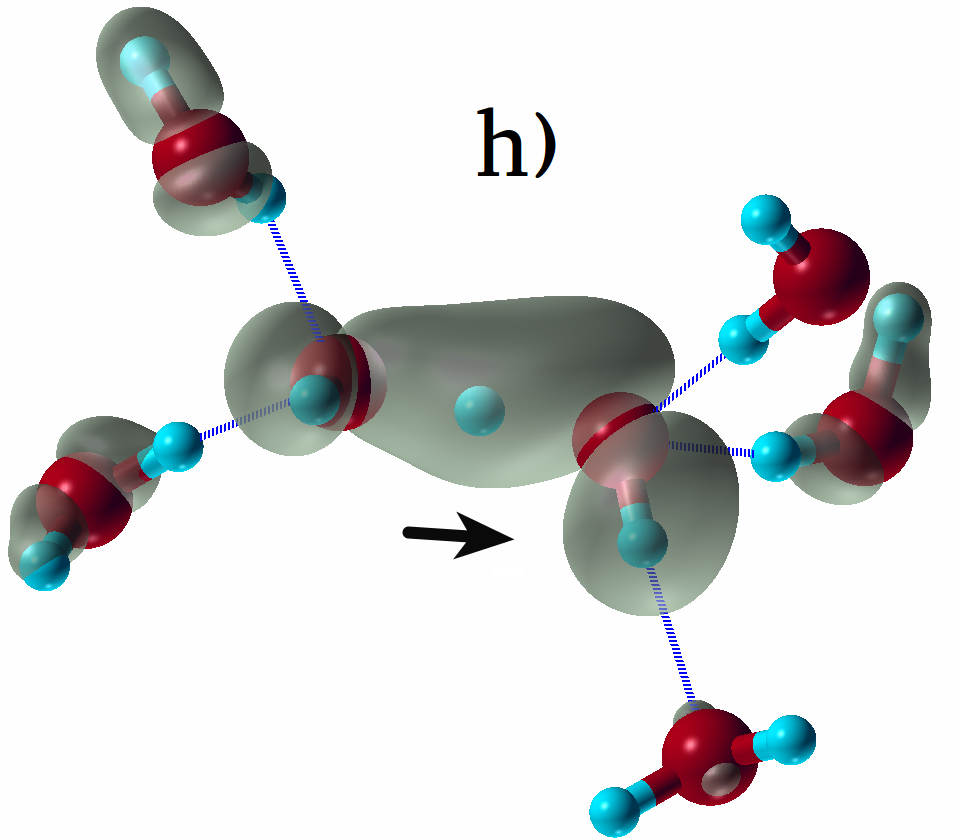} &\includegraphics[scale=0.09]{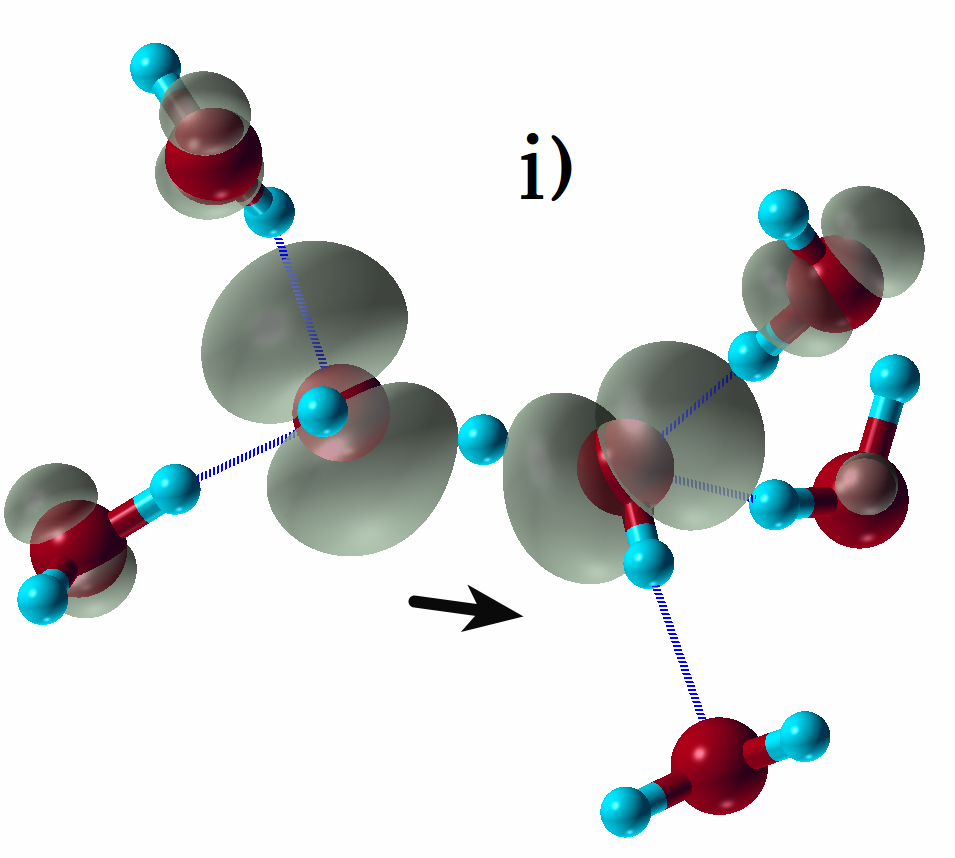}\\ \end{tabular}\\

          \hline
   \end{tabular}
   \begin{tabular}{|m{1.5cm}|c|}

          \hline
          \centering
          H$_3$O$^+$ & \begin{tabular}{ c@{\hspace{2.5cm}}c} $1b_2$ &$3a_1$\\ \end{tabular}\\
          \hline

          &\\
          Gas Phase & \begin{tabular}[c]{c@{\hspace{1.9cm}}c} \includegraphics[scale=0.025]{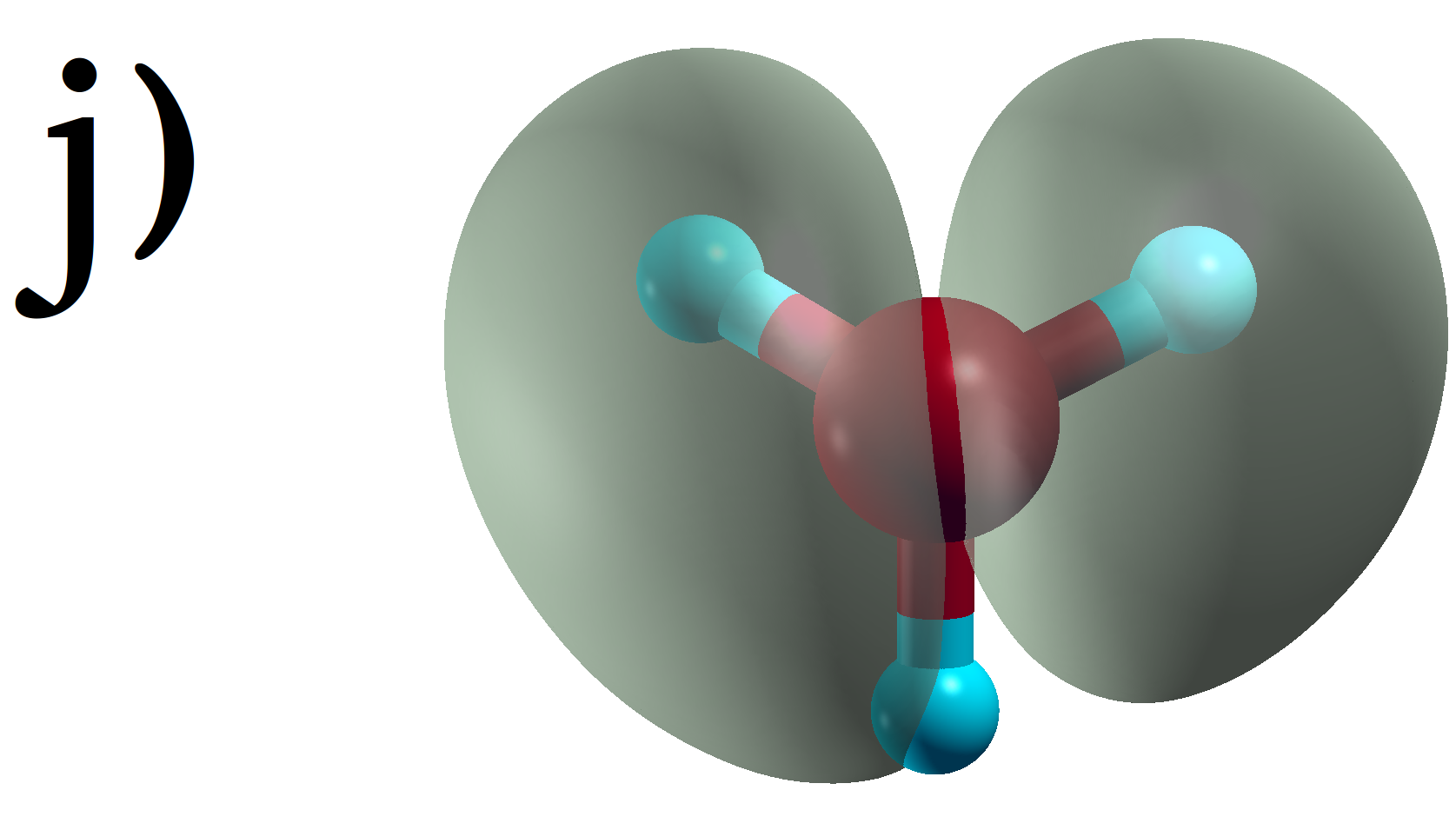} &\includegraphics[scale=0.025]{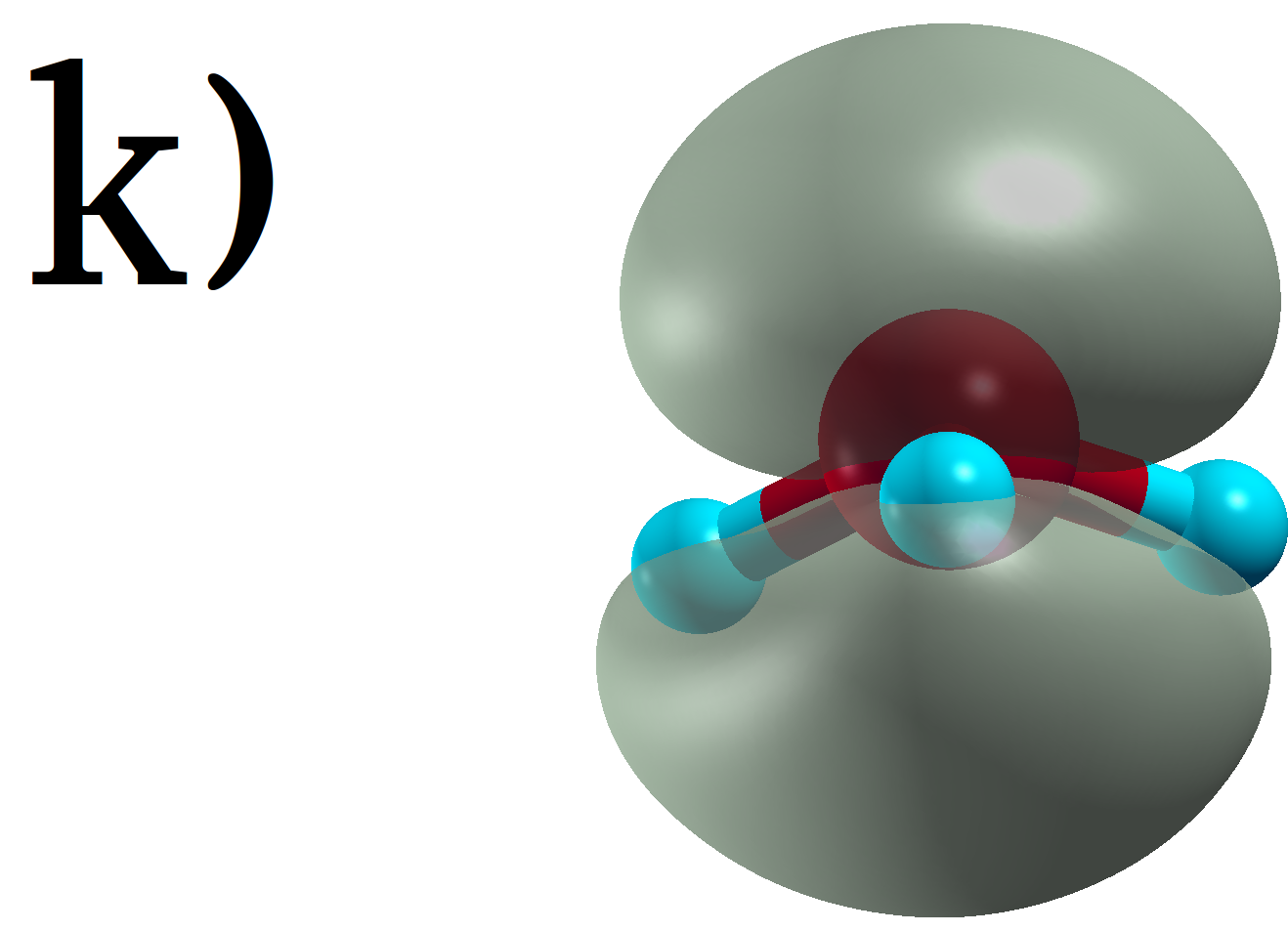}\\ \end{tabular}\\
         
          Liquid Solution: \newline No PT & \begin{tabular}[c]{cc} \includegraphics[scale=0.07]{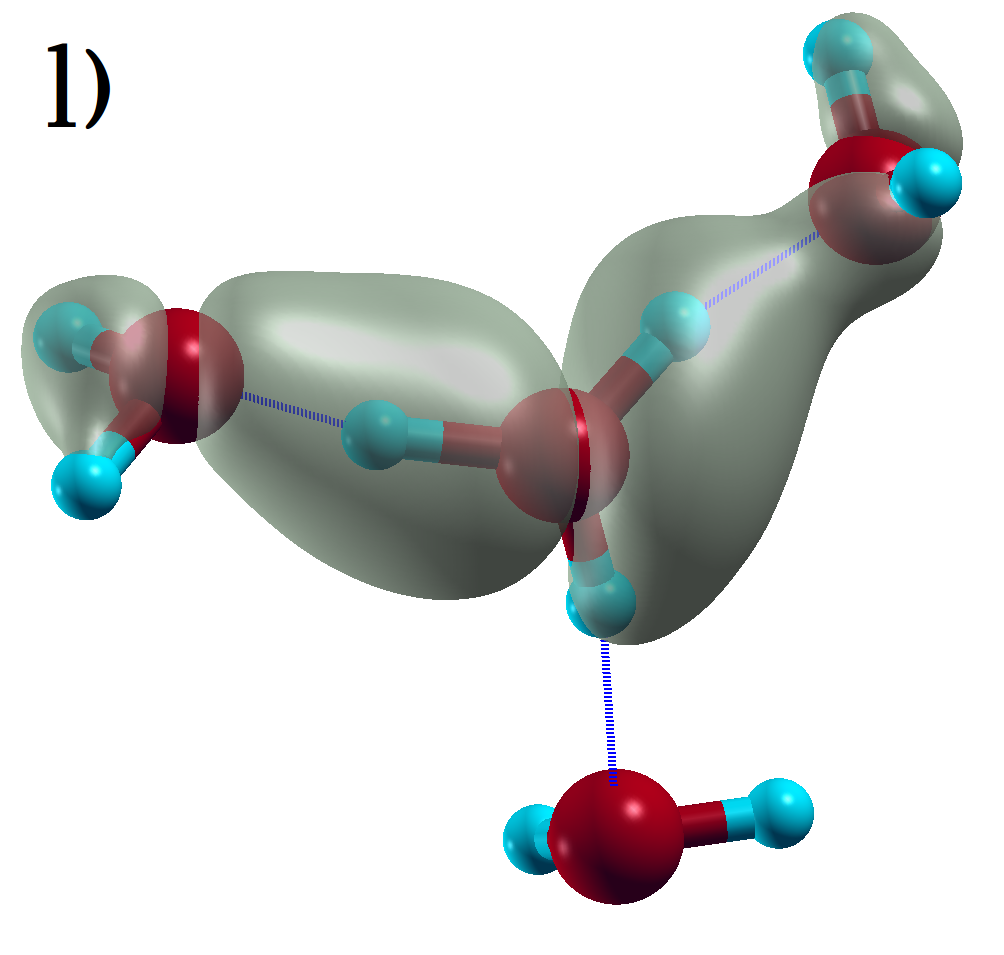} &\includegraphics[scale=0.07]{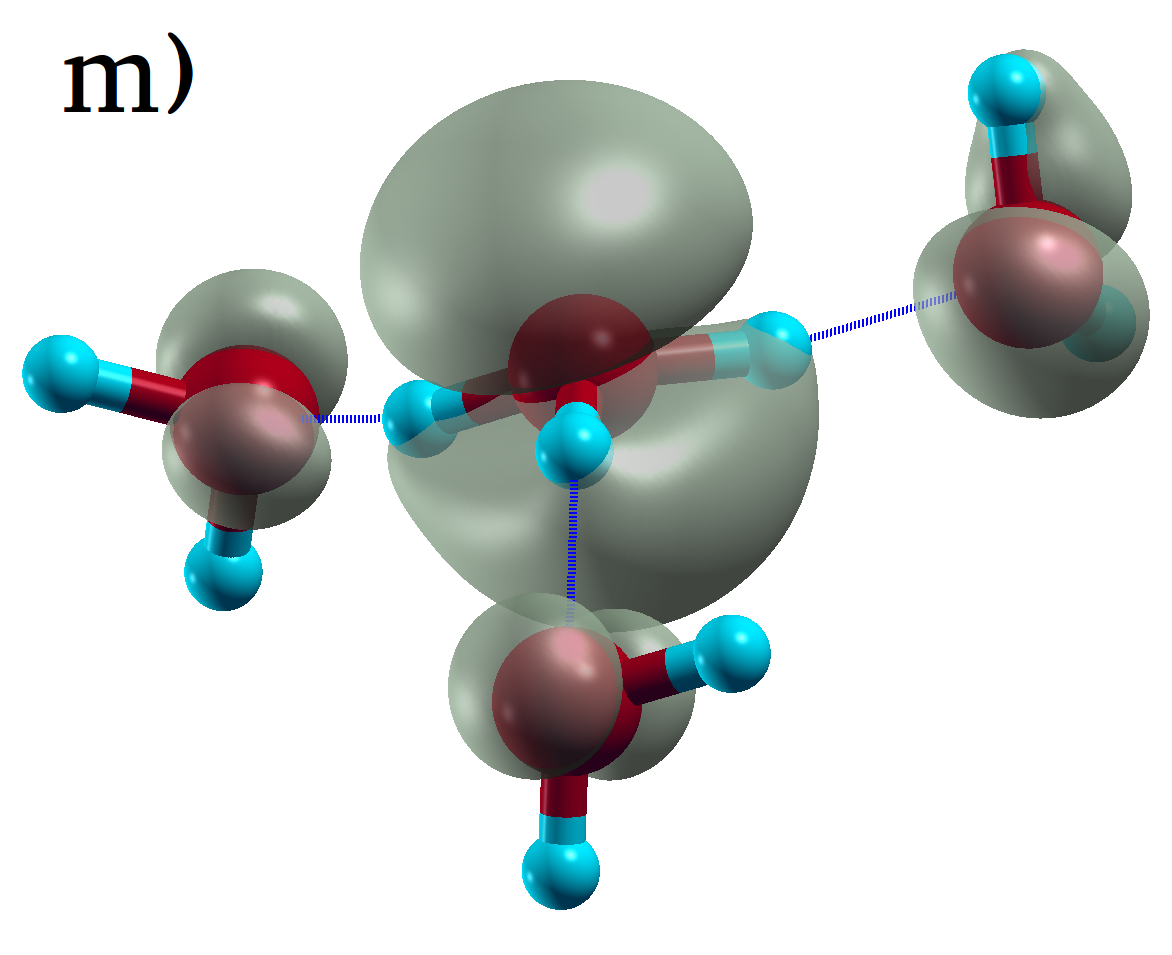}\\ \end{tabular}\\
         
          Liquid Solution: \newline PT& \begin{tabular}[c]{c@{\hspace{0.4cm}}c} \includegraphics[scale=0.06]{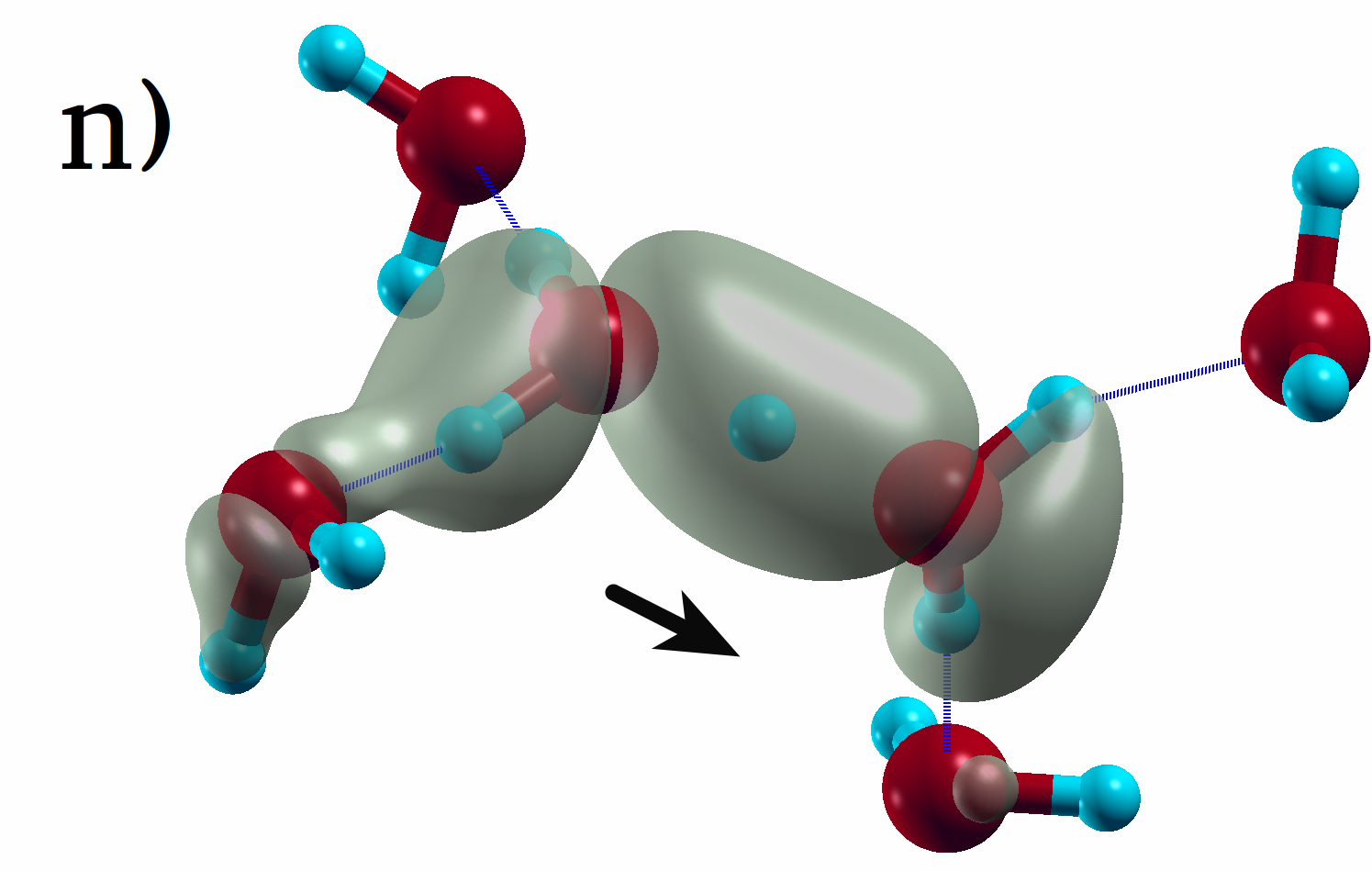} &\includegraphics[scale=0.06]{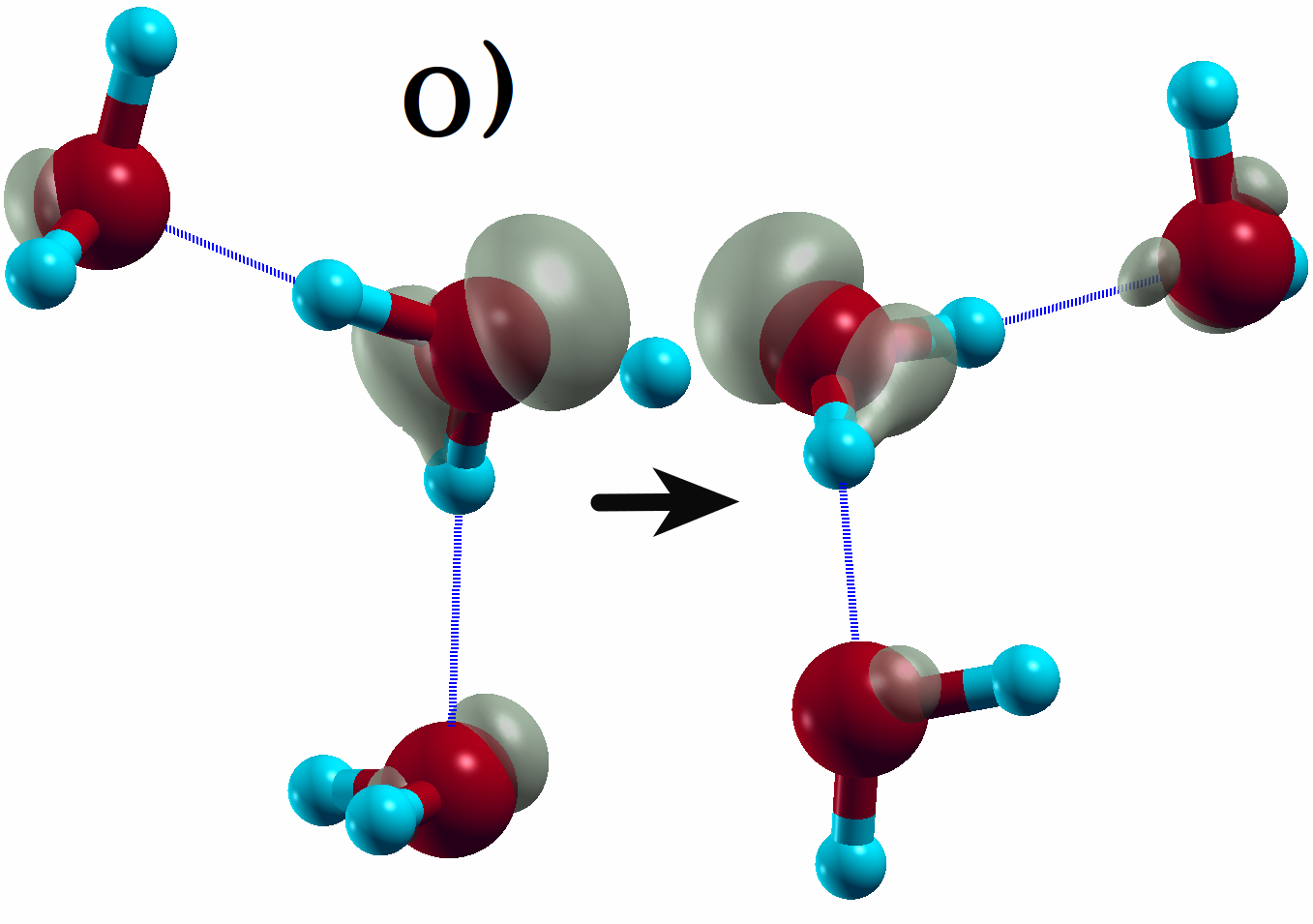}\\ \end{tabular}\\

          \hline
   \end{tabular}
   
   \caption{\label{fig:QW} (color online) QWs shown for gas phase (a)-(c)
      H$_2$O, (d)-(e) OH$^-$ and (j)-(k) H$_3$O$^+$.  Typical main feature $3a_1$ and $1b_1$ QWs
      displayed for hydrated OH$^-$ in liquid solution while (f)-(g) in a non-PT complex and (h)-(i)
      during a PT.  Similar $1b_2$ and $3a_1$ QWs displayed for hydrated H$_3$O$^+$ while (l)-(m) in
      the Eigen complex and (n)-(o) in the Zundel complex.  
   }
\end{figure}

In Fig.~\ref{fig:DOS_lDOS} we present our theoretical PES spectra of both ion solutions based on
quasiparticle density of states (QDOS). It can be seen that the overall PES spectra of both
hydronium and hydroxide solutions are dominated by three features belonging to bulk water, which have $1b_2$,
$3a_1$, and $1b_1$ characteristics with decreasing excitation energy as shown in Fig.~\ref{fig:QW}(a),
(b), and (c) respectively.  The IP is associated with the valence
electron excitation of solvated ion in water. Although embedded in the water solution PES spectra, it still
can be identified by the realspace projection of the QDOS onto the hydrated ions in the energy
range~\cite{defect_id}.  The resulting IP distributions of hydrated hydroxide and hydronium are also shown in the
shaded areas of Fig.~\ref{fig:DOS_lDOS}(a) and (b) respectively.  It can be seen that the main
feature of the IP of hydrated OH$^-$ is represented by a narrow distribution, whose peak
centered at 9.99 eV and close to the low energy edge of the $1b_1$ feature of bulk water. The
spectra signal is much less obvious in the $3a_1$ region and almost absent in the $1b_2$ region of
bulk water.  As far as the IP of hydrated H$_3$O$^+$  is concerned, the only prominent feature is
a broad peak centered at 19.01 eV close to the high energy edge of $1b_2$ feature of bulk water.
Much less signal is found in $3a_1$ region and almost no signal is found in $1b_1$ region of bulk
water.  Strikingly, current theory accurately reproduces the experimental
measurement~\cite{JACS_exp}, in which the IPs of hydrated hydroxide and hydronium are found to be
centered at $\sim$~9.2 eV and $\sim$~20 eV respectively.  The spectra distribution of solvated
hydronium is also found to be much broader than that of hydroxide in experiment~\cite{JACS_exp}.  In
stark contrast, the previous studies based on DFT calculations revealed that the IPs of solvated OH$^-$ and H$_3$O$^+$ were located at $\sim$~12 eV and $\sim$~5 eV respectively, which underestimated the
experimental value over 50\%~\cite{Tuckerman_LDA}. This is because DFT, as a ground state theory,
strongly underestimates the optical band gap properties. For a direct comparison with PES
experiments, the electron excitation process can be more appropriately described by the GW self-energy
approximation.  We want to stress here that the IPs for both solvated ions are
averaged over 100 configurations, weighted to include both PT~\cite{pt_def} and non-PT configurations, sampled
during the 25 ps equilibrium trajectory.  PT configurations account for $\sim10\%$ of the total OH$^-$
configurations and $\sim 16\%$ of the H$_3$O$^+$ configurations. 

The accurate quasiparticle predictions now enable more precise assignment of 
the IPs of hydrated ions. In Fig.~\ref{fig:QW}, we present the representative 
quasiparticle wavefunctions (QWs) for the main features, which are located close 
to $1b_1$ region of bulk water for hydrated OH$^-$ and close to $1b_2$ region
of bulk water for hydrated H$_3$O$^+$. The typical QWs of less prominent features
are also shown in Fig.~\ref{fig:QW}, whose energies are in the $3a_1$ region of 
bulk water for both ion solutions. 
For comparison, the three lowest IP energy states of the OH$^-$ and H$_3$O$^+$ monomer 
are also presented. The similarity of the electron excitation in ion solutions and that 
in gas phases indicates that the IPs should be attributed to molecular 
ionization, however, strongly perturbed by the solvation
structures of the surrounding water molecules.

   
\begin{figure}
   \captionsetup[subfigure]{labelformat=empty}
   \centering
   \subfloat[]{ \includegraphics[scale=0.539]{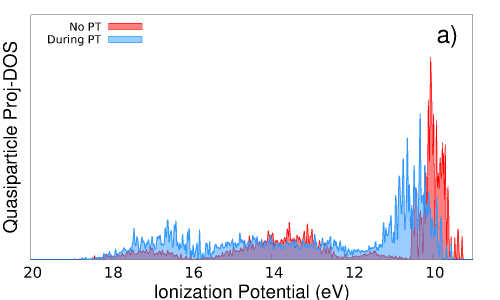}}
   \vspace{-0.5cm}
   \subfloat[]{ \includegraphics[scale=0.539]{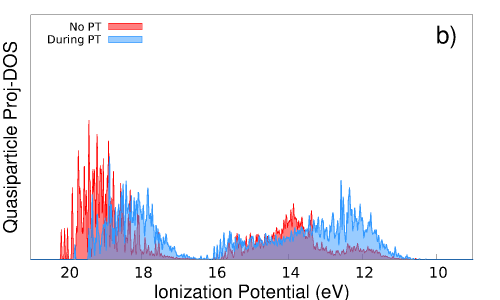}}
   \vspace{-0.5cm}
   \caption[]{ 
      (color online) Theoretical IP distributions for hydrated (a) OH$^-$ and (b) H$_3$O$^+$. The \textit{red} area indicates
      those configurations where the repective hydrated ion is not experiencing a PT ($\delta > 0.6$ \AA~\cite{delta_def}) 
      and the \textit{blue} area indicates those configurations currently experiencing a PT ($\delta \sim 0$ \AA). 
   }
   \label{fig:lDOS_compare}
\end{figure}

Clearly, the main feature of the IP in hydrated hydroxide at 9.99 eV should be 
assigned to the first valence electron excitation of OH$^-$ monomer.
This is due to the resemblance of the electron excitations between  
the gas phase and the liquid solution. In the aqueous solution, 
the typical QW of the main feature is a well localized hydroxide defect 
state of clear lone pair character as shown in Fig.~\ref{fig:QW}(g) and (i). 
It originates from two degenerate $1\pi$ bonds of the OH$^-$ monomer.
The degeneracy is broken by the disordered liquid structure and results in
an excitation distribution instead of one single ionization energy. 
It can be expected that the spectra distribution is largely dependent on
the solvation structure and will be different with or without PT.
In the absence of PT, current AIMD simulation finds that solvated hydroxide
adopts the most stable configuration in such a way that OH$^-$ accepts
four H-bonds with the possibility of donating one~\cite{Tuckerman_review_2010}.
The four water molecules donating H-bond are approximately in one plane.
Consistently this lone pair QW is mainly localized on the OH$^-$ itself,
however, with a significant weight on the water molecules within the 
first solvation shell. Thus without PT, the IP is mainly 
affected by the fluctuating first-shell solvation structure 
embedded in the H-bond network of liquid water.
During PT, the OH$^-$ exchanges one proton with neighboring water molecule.
Our current simulation indicates that the PT mechanism is consistent with
the {\it dynamical hypercoordination} scenario proposed by Tuckerman {\it et al}~\cite{Tuckerman_review_2010}.
The structural diffusion is initiated by breaking one of the four
accepting H-bonds and followed by a proton exchange along the 
shortest H-bond. Intriguingly, the structural diffusion is accompanied by a nontrivial
change in the electronic state. The lone pair QW 
is now localized on both proton donating and proton receiving
molecules as shown in Fig.~\ref{fig:QW}(i) instead of its main localization 
on OH$^-$ only before the PT in Fig.~\ref{fig:QW}(g).
The delocalized QW facilitates its hybridization with surrounding water molecules.
As a result, the IP exhibits a blue shift 
towards the main feature of the bulk water spectra accompanied by
a broadened distribution as illustrated in Fig.~\ref{fig:lDOS_compare}(a).

In a similar scenario, we assign the main feature of the IP distribution
of hydrated hydronium to the second valence excitation of 
two degenerate $p$ orbitals of H$_3$O$^+$ monomer. The degeneracy is again
broken by the disordered molecular environment.
Without PT, the excited QW has a similar
characteristic of $1b_2$ of liquid water and is 
mainly localized on the so-called Eigen complex (H$_3$O$^+$) as shown in Fig.~\ref{fig:QW}(l).
In particular, a large orbital amplitude is also found on the three
water molecules along the direction of donated H-bonds by H$_3$O$^+$.
The QW then decays rapidly to zero in the second solvation shell and beyond.
As a result, the excitation is a hydronium defect state with $1b_2$ character
that is localized on the H$_9$O$_4^+$ complex, which is 
also the so-called {\it strongly solvated Eigen cation} in the literature~\cite{News}. 
In the process of PT, the current simulation is consistent with the widely accepted 
Grottuss diffusion mechanism~\cite{Tuckerman_review_2010}. The proton exchange takes place
by the interconversion between Eigen and Zundal (H$_5$O$_2^+$) complex~\cite{Tuckerman_review_2010}.
Not surprisingly, the locality of the QW also swifts from the 
Eigen to the Zundal complex during PT as shown in Fig.~\ref{fig:QW}(l) and (n)
respectively. Interestingly, a significant weight is also found to be centered on the
transferring proton connecting both the proton receiving and donating structures, which
gives a unique signal for PT in the electronic structure.
Similar to the PT process of hydrated OH$^-$, the delocalized H$_3$O$^+$ defect state
is much more easily to be hybridized with solvent H$_2$O molecules.
As a result, in Fig.~\ref{fig:lDOS_compare}(b) 
we observe a red shift of the IPs into the $1b_2$ region of bulk
water which gets broadened simultaneously.

Interestingly, a comparison of the IPs 
between these two ion solutions reveals that the main feature of 
hydrated H$_3$O$^+$ has a much broader distribution than that of OH$^-$.
We attribute it to the difference in orbital characters.
For the hydrated H$_3$O$^+$ excitation with $1b_2$ characteristic, 
the covalent orbital on an OH bond is easily perturbed by the H-bond network of liquid water. 
On the other hand, the excitation of hydrated OH$^-$ is of lone pair character
and only centered on oxygen atom which will be much less effected by
the embedded H-bond network. This is also consistent with the 
observation that $1b_1$ is already narrowed than $1b_2$ peak in liquid water~\cite{water_DFT}.

\begin{figure}
   \captionsetup[subfigure]{labelformat=empty}
   \subfloat[]{ \includegraphics[scale=0.33]{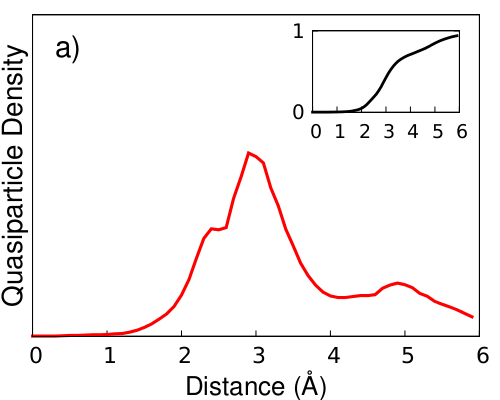}}
   \subfloat[]{ \includegraphics[scale=0.33]{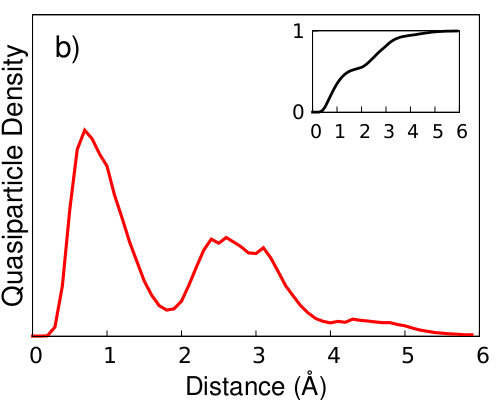}}
   \vspace{-0.6cm}
   \subfloat[]{ \includegraphics[scale=0.33]{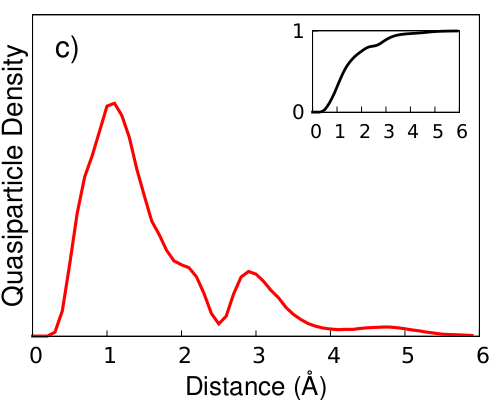}}
   \subfloat[]{ \includegraphics[scale=0.33]{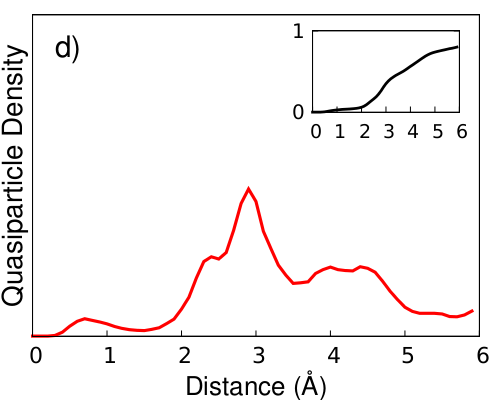}}
   \vspace{-0.6cm}
   \caption[]{
      Typical QW densities (arb. u.) for hydrated OH$^-$ characteristic states, (a) $3a_1$ and (b)
      $1b_1$, and  hydrated H$_3$O$^+$ characteristic states, (c) $1b_2$ and (d) $3a_1$, as
      a function of distance from O$^*$\cite{defect_id}. Inserts show the total integrated density. 
   
   }
   \label{fig:KS_proj}
\end{figure}

Besides the main features, the excitations of hydrated ions are also found
in Fig.~\ref{fig:DOS_lDOS}(a) and (b) within the $3a_1$ region of bulk water 
for both solutions.
However the signals are much weaker and broadly distributed. 
Nevertheless, the IPs of hydrated ions should still be assigned to the molecular
excitation of the same $3a_1$ symmetry, which are the second and first excitation
of OH$^-$ and H$_3$O$^+$ monomers respectively.
In Fig.~\ref{fig:KS_proj}(a) and (d), we present typical QW densities
of $3a_1$ excitations as a function of distance from the solvated OH$^-$ and H$_3$O$^+$ respectively.
For comparison, the same quantities of the main feature   
are also shown in Fig.~\ref{fig:KS_proj}(b) and (c). 
Clearly, in both ion solutions, the less prominent IPs of $3a_1$ character
originate from relatively delocalized QWs, in which a large orbital amplitude 
falls into the second solvation shell and beyond. 
On the contrary, QWs of the main features of both hydrated ions are
strongly localized defect states within the first solvation shell.
As a result, the integrated charge approaching to one electron rapidly 
with the increased distance away from the ions.
The delocalized QWs indicate a stronger hybridization with the water solution
and result in the observed broader and weaker ion excitations.

Finally, we draw our attention to the energy range where the least IP signals have been
found. They are the $1b_2$ and $1b_1$ region of bulk water for hydrated 
OH$^-$ and H$_3$O$^+$ respectively. Again, this can be understood by the symmetry
of electron excitation of both ions at the monomer level.
The electronic configurations of OH$^-$ and H$_3$O$^+$ ions are intrinsically different from 
that  of a single water molecule. For OH$^-$ monomer, the three lowest allowed valence excitation 
are the two degenerated excitation of $1b_1$ character, followed by one of $3a_1$ character,
in which the $1b_2$-like orbital in water is not allowed by symmetry.
On the other hand, the three lowest valence excitation in H$_3$O$^+$ ion are
one with $3a_1$ character, followed by two degenerate states with $1b_2$ character,
in which the lone pair $1b_1$-like orbital in water is forbidden.
These symmetry restrictions are also reflected in ion solutions and result in the 
absence of electron excitation for hydrated ions in the above energy range. 

In conclusion, the IP distributions of hydrated OH$^-$ and H$_3$O$^+$
are studied by accurate quasiparticle theory. The excitations of solvated ions
can be assigned to molecular electron excitation, however, strongly perturbed 
by the solvation structures. Although the main features of IPs are 
determined by their stable configurations, the proton transfer does 
introduce a change of  position and distributions of electron excitation in both hydrated ions.
We suggest that the excitation change due to PT can be detected by isotope effect in future 
PES measurements performed on solvated H$_3$O$^+$ and D$_3$O$^+$. 
Because of the quantum nuclear effect lowers the barrier of proton transfer~\cite{Marx_Nature}, the hydrated H$_3$O$^+$
will have a more broad and shifted IP distribution than that of D$_3$O$^+$. 
Finally we comment that the leftover mismatch between experiment and theory could be further
reduced by more accurate liquid structures considering both dispersion force~\cite{Zhang_JCTC, Biswajit_PRL, Rob_PNAS} and 
self-interaction error correction~\cite{Cui_Cl,Cui_PBE0_water} and by
including frequency dependence of excitation beyond static GW~\cite{Kang_PRB, Lu_PRL_2008}. 

XW thanks valuable discussion with Roberto Car. This work is supported by 
U.S. Department of Energy under Grant No. DE-SC0008726. 
Computational support is provided by the National Energy 
Research Scientific Computing Center.


\begin{thebibliography}{50}

\bibitem{Tuckerman_review_2010}
D. Marx, A. Chandra, M. K. Tuckerman, Chem. Rev. {\bf 110}, 2174 (2010).

\bibitem{Tuckerman_ACR}
M. Tuckerman, A. Chandra, and D. Marx, Acc. Chem. Res. {\bf 39}, 151 (2006)

\bibitem{Parrinello_PNAS_2011}
A. Hassanali1, M. K. Prakash, H. Eshet, and M. Parrinello, Proc. Natl. Acad. Sci. {\bf 108}, 20410 (2011).


\bibitem{Hynes_Nature}
J. T. Hynes, Nature {\bf 397}, 565 (1999).

\bibitem{Asthagiri_PNAS}
D. Asthagiri, L. Pratt, J.D. Kress, and M. Gomez, Proc. Natl. Acad. Sci. {\bf 101}, 19 (2004).

\bibitem{JACS_exp}
B. Winter, M. Faubel, I. Hertel, C. Pettenkofer, S. Bradforth, B. Jagoda-Cwiklik, L. Cwiklik, and P. Jungwirth, J. Am. Chem. Soc. {\bf 128}, 3864 (2006).

\bibitem{Chandra_PRL_2007}
A. Chandra, M. Tuckerman, and D. Marx, Phys. Rev. Lett. {\bf 99}, 145901 (2007).

\bibitem{Winter_JACS_2005}
B. Winter, R. Weber, I. V. Hertel, M. Faubel, P. Jungwirth, E. C. Brown, and S. E. Bradforth, 
J. Am. Chem. Soc. {\bf 127}, 7203 (2005).

\bibitem{Tuckerman_LDA}
M. Tuckerman, K Laasonen, and M Parrinello, J. Chem. Phys. {\bf 103}, 150 (1995).

\bibitem{AIMD}
R. Car, and M. Parrinello, Phys. Rev. Lett, {\bf 55}, 2471 (1985).

\bibitem{Patrick_PRB}
P. Sit, C. Bellin, B. Barbiellini, D. Testemale, J. -L. Hazamann, T. Buslaps, N. Marzari, and A. Shukla, 
Phys. Rev. B {\bf 76}, 245413 (2007).

\bibitem{Hedin}
L. Hedin, Phys. Rev. {\bf 139}, A796 (1965).

\bibitem{Onida}
G. Onida, L. Reining, and A. Rubio, Rev. Mod. Phys. {\bf 74}, 601 (2002).

\bibitem{Model_GW_1988}
M. S. Hybertsen and S. G. Louie, Phys. Rev. B {\bf 37}, 2733 (1988).

\bibitem{Wu_PRB_2009}
X. Wu, A. Selloni, and R. Car, Phys. Rev. B {\bf 79}, 085102 (2009).

\bibitem{Wei_PRL}
W. Chen, X. Wu, and R. Car, Phys. Rev. Lett. {\bf 105}, 017802 (2010).

\bibitem{Lingzhu_PRB_2012}
L. Kong, X. Wu, and R. Car, Phys. Rev. B {\bf 86}, 134203 (2012).

\bibitem{MLWF}
N. Marzari, A. A. Mostofi, J. R. Yates, I. Souza, and D. Vanderbilt, Rev. Mod. Phys.
{\bf 84}, 1419 (2012).


\bibitem{Zhaofeng_Li}
Z. Li, Ph.D. thesis, Physics Department, Princeton University, 2012.

\bibitem{PBE}
J. P. Perdew,K. Burke, and M. Ernzerhof, Phys. Rev. Lett. {\bf 77}, 3865 (1996).

\bibitem{QuantumEspresso}
   P. Giannozzi, P. Baroni, N. Bonini, M. Calandra, R. Car, C. Cavazzoni, D. Ceresoli, G. L. Chiarrotti, M. Cococcioni, and I. Dabo, J. Phys. Condens. Matter. {\bf 21}, 395502 (2005) 


\bibitem{defect_id}
   Individual IPs were calculated by projecting
   the QWs on to a real-space sphere of 1.5 \AA\cite{Tuckerman_LDA} \space centered on the O$^*$ atom 
   while in a stable complex and centered on the transferring proton during PT~\cite{pt_def}. 
   The IPs of hydrated ions are amplified by a factor of 10 in order to increase their visibility.
   The identity of both the OH$^-$ and the H$_3$O$^+$ are determined by counting the number of H$^+$
   inside a covalent radius of 1.17 \AA \space for each oxygen atom in a given configuration.
   Those oxygens containing one or 3 H atoms within their covalent radius are referred to as O$^*$
   and determine both OH$^-$ and H$_3$O$^+$ respectively. 

 
\bibitem{pt_def}
   Configurations are considered to be in a Proton transfer event when a single H$^+$ could not be associated 
   with any oxygen molecules, in accordance with Ref.~\cite{defect_id}.

\bibitem{News}
B. Kirchner, Chem. Phys. Chem. {\bf 8}, 41 (2007).

\bibitem{water_DFT}
D. Prenergast, J. Grossman, and G. Galli, J. Chem. Phys. {\bf 123}, 014501 (2005)

\bibitem{delta_def} 
   $\delta$ is defined as $\delta = \left|R_{{\rm O}^{*}{\rm H}} - R_{\rm O_{w}H}\right|$, where $R_{\rm O^{*}H}$ and
   $R_{\rm O_{w}H}$ are the distances between a shared proton and O$^*$~\cite{defect_id} or a surrounding
   water molecule's oxygen, respectively. The minimum $\delta$ value for a particular O$^*$ is
   considered most likely to produce a PT, and $\delta = 0$ indicates that the shared proton is
   halfway between two molecules~\cite{Tuckerman_ACR}.

\bibitem{Marx_Nature}
D. Marx, M. Tuckerman, J. Hutter, and M. Parrinello, Nature {\bf 397}, 601 (1999).


\bibitem{Zhang_JCTC}
C. Zhang, and J. Wu, G. Galli, and F. Gygi, J. Chem. Theory Comp. {\bf 7}, 3054 (2011).


\bibitem{Biswajit_PRL}
B. Santra, J. Klime\v{s}, D. Alf\'{e}, A. Tkatchenko, Ben Slater, A. Michaelides, R. Car, and M. Scheffler,
Phys. Rev. Lett. {\bf 107}, 185701 (2011).

\bibitem{Rob_PNAS}
R. A. DiStasio, O. Anatole von Lilienfeld, and A. Tkatchenko, Proc. Natl. Acad. Sci. {\bf 109}, 14791 (2012).

\bibitem{Cui_Cl}
C. Zhang, T. A. Pham, F. Gygi, and G. Galli, J. Chem. Phys. {\bf 109}, 14791 (2013).

\bibitem{Cui_PBE0_water}
C. Zhang, D. Donadio, F. Gygi, and G. Galli
J. Chem. Theory Comput. {\bf 7}, 1443 (2011).

\bibitem{Kang_PRB}
W. Kang and M. S. Hybertsen, Phys. Rev. B {\bf 82}, 195108 (2010).

\bibitem{Lu_PRL_2008}
D. Lu, F. Gygi, and G. Galli, Phys. Rev. Lett., {\bf 100}, 147601 (2008).

\bibitem{Comment_GW}
The static GW is found to slightly overestimate the IP~\cite{Onida, Kang_PRB}, which can be
corrected by frequency dependence in GW method.



\end{thebibliography}
\end{document}